AIAA-2007-0590

# Wireless Transfer of Electricity in Outer Space*

Alexander Bolonkin
C&R, 1310 Avenue R, #F-6, Brooklyn, NY 11229, USA
T/F 718-339-4563, aBolonkin@juno.com, http://Bolonkin.narod.ru

**Abstract**
Author offers conclusions from his research of a revolutionary new idea - transferring electric energy in the hard vacuum of outer space wirelessly, using a plasma power cord as an electric cable (wire). He shows that a certain minimal electric currency creates a compressed force that supports the plasma cable in the compacted form. A large energy can be transferred hundreds of millions of kilometers by this method. The required mass of the plasma cable is only hundreds of grams. He computed the macroprojects: transference of hundreds kilowatts of energy to Earth's Space Station, transferring energy to the Moon or back, transferring energy to a spaceship at distance 100 million of kilometers, the transfer energy to Mars when one is located at opposed side of the distant Sun, transfer colossal energy from one of Earth's continents to another continent (for example, between Europe-USA) wirelessly—using Earth's ionosphere as cable, using Earth as gigantic storage of electric energy, using the plasma ring as huge MagSail for moving of spaceships. He also demonstrates that electric currency in a plasma cord can accelerate or brake spacecraft and space apparatus.
--------------------------

## Introduction

   The production, storage, and transference of large amounts of electric energy is an enormous problem for humanity, especially of energy transfer in outer space (vacuum). These spheres of industry are search for, and badly need revolutionary ideas. If in production of energy, space launch and flight we have new ideas (see [1]-[17]), it is not revolutionary ideas in transferring and storage energy except the work [4].
 However, if we solve the problem of transferring energy in outer space, then we solve the many problems of manned and unmanned space flight. For example, spaceships can move long distances by using efficient electric engines, orbiting satellites can operate unlimited time periods without entry to Earth's atmosphere, communication satellites can transfer a strong signal directly to customers, the International Space Station's users can conduct many practical experiments and the global space industry can produce new materials.  In the future, Moon and Mars outposts can better exploration the celestial bodies on which they are placed at considerable expense.
   Other important Earth mega-problem is efficient transfer of electric energy long distances (intra-national, international, intercontinental). The consumption of electric energy strongly depends on time (day or night), weather (hot or cold), from season (summer or winter). But electric station can operate most efficiently in a permanent base-load generation regime. We need to transfer the energy a far distance to any region that requires a supply in any given moment or in the special hydro-accumulator stations. Nowadays, a lot of loss occurs from such energy transformation. One solution for this macro-problem is to transfer energy from Europe to the USA during nighttime in Europe and from the USA to Europe when it is night in the USA. Another solution is efficient energy storage, which allows people the option to save electric energy.
   The storage of a big electric energy can help to solve the problem of cheap space launch. The problem of an acceleration of a spaceship can be solved by use of a new linear electrostatic engine suggested in [5].  However, the cheap cable space launch offered by author [4] requires use of gigantic energy in short time period. (It is inevitable for any launch method because we must accelerate big masses to the very high speed - 8 ÷11 km/s). But it is impossible to turn off whole state and connect all electric station to one customer. The offered electric energy storage can help solving this mega-problem for humanity [5]-[17].

## Offered Innovations and Brief Descriptions

The author offers the series of innovations that may solve the many macro-problems of transportation energy in space, and the transportation and storage energy within Earth's biosphere. Below are some of them.

1) transfer of electrical energy in outer space using the conductive cord from plasma. Author solved the main problem - how to keep plasma cord in compressed form. He developed theory of space electric transference, made computations that show the possibility of realization for these ideas with existing technology. The electric energy may be transferred in hundreds millions of kilometers in space (include Moon and Mars).
2) method of construction for space electric lines and electric devices.
3) method of utilization of the plasma cable electric energy.
4) a new very perspective gigantic plasma MagSail for use in outer space as well as a new method for connection the plasma MagSail to spaceship.
5) a new method of projecting a big electric energy through the Earth's ionosphere.
6) a new method for storage of a big electric energy used Earth as a gigantic spherical condenser.
7) a new propulsion system used longitudinal (cable axis) force of electric currency.

Below are some succinct descriptions of some constructions made possible by these revolutionary ideas.

**1. Transferring electric energy in Space**. The electric source (generator, station) is connected to a space apparatus, space station or other planet by two artificial rare plasma cables (Fig.1a). These cables can be created by plasma beam [7] sent from the space station or other apparatus.

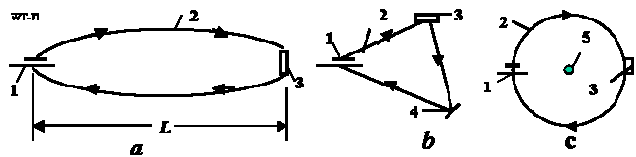

**Fig.1.** Long distance plasma transfer electric energy in outer space. *a* - Parallel plasma transfer, *b* - Triangle plasma transfer, *c* - circle plasma transfer. Notations: 1 - current source (generator), 2 - plasma wire (cable), 3 - spaceship, orbital station or other energy addresses, 4 - plasma reflector, 5 - central body.

The plasma beam may be also made the space apparatus from an ultra-cold plasma [7] when apparatus starting from the source or a special rocket. The plasma cable is self-supported in cable form by magnetic field created by electric currency in plasma cable because the magnetic field produces a magnetic pressure opposed to a gas dynamic plasma pressure (teta-pinch)(Fig. 2). The plasma has a good conductivity (equal silver and more) and the plasma cable can have a very big cross-section area (up thousands of square meter). The plasma conductivity does not depend on its density. That way the plasma cable has a no large resistance although the length of plasma cable is hundreds millions of kilometers. The needed minimum electric currency from parameters of a plasma cable researched in theoretical section of this article.

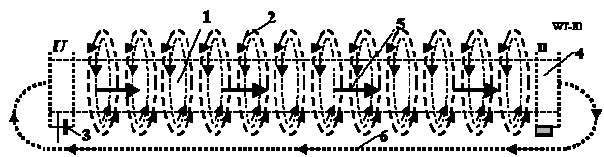

**Fig.2.** A plasma cable supported by self-magnetic field. Notations: 1 -plasma cable, 2 - compressing magnetic field, 3 - electric source, 4 - electric receiver, 5 - electric currency, 6 - back plasma line.

The parallel cables having opposed currency repels one from other (Fig.1a). They also can be separated by a special plasma reflector as it shown in figs. 1b, 2c. The electric cable of the plasma transfer can be made circular (Fig. 1c). The radial compressed magnetic force from a circle currency may be balanced a small rotation of the plasma cable (see theoretical section). The circle form is comfortable for building the big plasma cable lines for spaceship not having equipment for building own electric lines or before a space launch. We build small circle and gradually increase the diameter up to requisite value (or up spaceship). The spaceship connects to line in suitable point. Change the diameter and direction of plasma circle we support the energy of space apparatus. At any time the spaceship can disconnect from line and circle line can exist without user.

The electric tension (voltage) in a plasma cable is made two nets in issue electric station (electric generator) [7]-[8]. The author offers two methods for extraction of energy from the electric cable (Fig.3) by customer (energy addresses). The plasma cable currency has two flows: electrons (negative) flow and opposed ions (positive) flow in one cable. These flows create an electric current. (It may be instances when ion flow is stopped and current is transferred only the electron flow as in a solid metal or by the ions flow as in a liquid electrolyte. It may be the case when electron-ion flow is moved in same direction but electrons and ions have different speeds). In the first method the two nets create the opposed electrostatic field in plasma cable (resistance in the electric cable [7]-[8]) (figs.1, 3b). This apparatus resistance utilizes the electric energy for the spaceship or space station. In the second method the charged particles are collected a set of thin films (Fig. 3a) and emit (after utilization in apparatus) back into continued plasma cable (Fig.3a)(see also [7]-[8]).

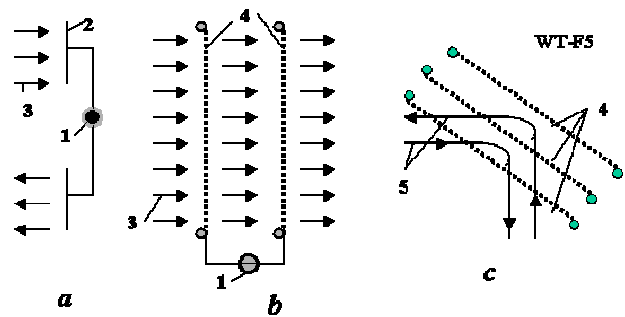

**Fig.3.** Getting the plasma currency energy from plasma cable. *a* - getting by two thin conducting films; *b* - getting two nets which brake the electric current flux; *c* - plasma reflector. Notations: 1 - spaceship or space station, 2 - set (films) for collect (emit) the charged particles, 3 - plasma cable, 4 - electrostatic nets.

Fig. 3c presents the plasma beam reflector [7]-[8]. That has three charged nets. The first and second nets reflect (for example) positive particles, the second and third nets reflected the particles having an opposed charge.

**2. Transmitting of the electric energy to satellite, Earth's Space Station, or Moon.** The suggested method can be applied for transferring of electric energy to space satellites and the Moon. For transmitting energy from Earth we need a space tower of height up 100 km, because the Earth's atmosphere will wash out the plasma cable or we must spend a lot of energy for plasma support. The design of solid, inflatable, and kinetic space towers are revealed in [4],[13]-[14],[16].

It is possible this problem may be solved with an air balloon located at 30-45 km altitude and connected by conventional wire with Earth's electric generator. Further computation can make clear this possibility.

If transferring valid for one occasion only, that can be made as the straight plasma cable 4 (Fig. 4). For multi-applications the elliptic closed-loop plasma cable 6 is better. For permanent transmission the Earth must have a minimum two space towers (Fig.4). Many solar panels can be located on Moon and Moon can transfer energy to Earth.

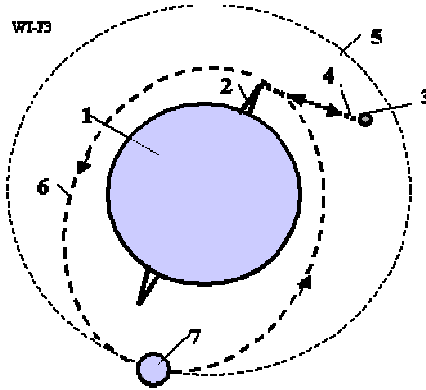

**Fig.4.** Transferring electric energy from Earth to satellite, Earth's International Space Station or to Moon (or back) by plasma cable. Notations: 1 - Earth, 2 - Earth's tower 100 km or more, 3 - satellite or Moon, 4 - plasma cable, 5 - Moon orbit, 6 - plasma cable to Moon, 7 - Moon.

**3. Transferring energy to Mars**. The offered method may be applied for transferring energy to Mars including the case when Mars may be located in opposed place of Sun (Fig. 5). The computed macroproject is in Macroprojects section.

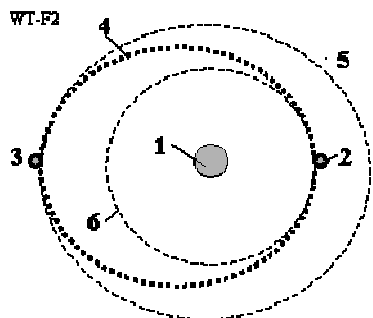

**Fig.5.** Transferring of electric energy from Earth to Mars located in opposed side of Sun. Notations: 1 - Sun, 2 - Earth, 3 - Mars, 4 - circle plasma cable.

**4. Plasma AB Magnetic Sail**. Very interesting idea to build a gigantic plasma circle and use it as a Magnetic Sail (Fig. 6) harnessing the Solar Wind. The computations show (see section "Macroproject") that the electric resistance of plasma cable is small and the big magnetic energy of plasma circle is enough for existence of a working circle in some years without external support. The connection of spaceship to plasma is also very easy. The space ship create own magnetic field and attracts to MagSail circle (if spacecraft is located behind the ring) or repels from MagSail circle (if spaceship located ahead of the ring). The control (turning of plasma circle) is also relatively easy. By moving the spaceship along the circle plate, we then create the asymmetric force and turning the circle. This easy method of building the any size plasma circle was discussed above.

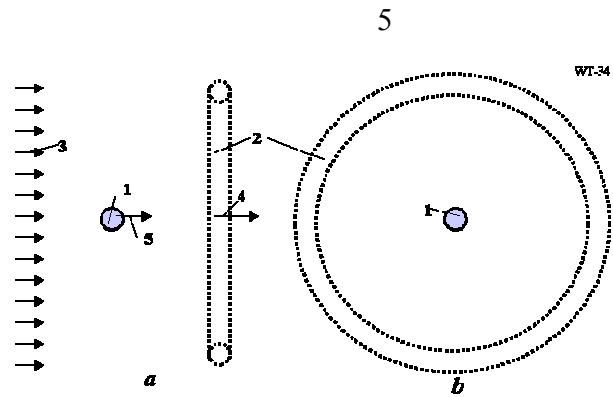

**Fig.6**. Plasma AB-MarSail. Notations: 1 - spaceship, 2 - plasma ring (circle), 3 - Solar wind, 4 - MagSail thrust, 5 - magnetic force of spaceship.

**5. Wireless transferring of electric energy in Earth**. It is interesting the idea of energy transfer from one Earth continent to another continent without wires. As it is known the resistance of infinity (very large) conducting medium does not depend from distance. That is widely using in communication. The sender and receiver are connected by only one wire, the other wire is Earth. The author offers to use the Earth's ionosphere as the second plasma cable. It is known the Earth has the first ionosphere layer $E$ at altitude about 100 km (Fig. 7). The concentration of electrons in this layer reaches $5 \times 10^4$ $1/cm^3$ in daytime and $3.1 \times 10^3$ $1/cm^3$ at night (Fig. 7). This layer can be used as a conducting medium for transfer electric energy and communication in any point of the Earth. We need minimum two space 100 km. towers (Fig. 8). The cheap optimal inflatable, kinetic, and solid space towers are offered and researched by author in [4], [6], [7], [16]. Additional innovations are a large inflatable conducting balloon at the end of the tower and big conducting plates in a sea (ocean) that would dramatically decrease the contact resistance of the electric system and conducting medium.

Theory and computation of these ideas are presented in Macroprojects section.

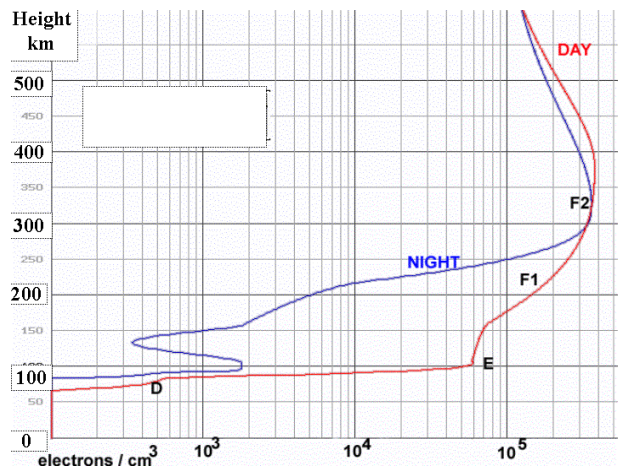

**Fig.7**. Consentration/$cm^3$ of electrons (= ions) in Earth's atmosphere in the day and night time in the D, E, F1, and F2 layers of ionosphere.

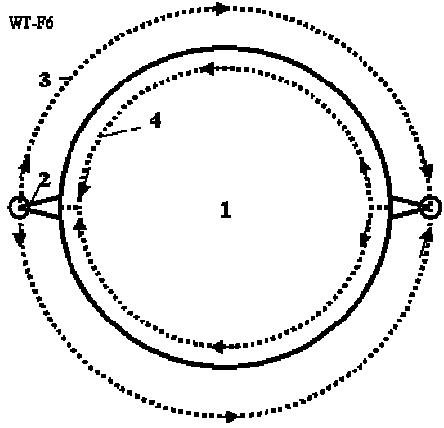

**Fig.8**. Using the ionosphere as conducting medium for transferring a huge electric energy between continents and as a large storage of the electric energy. Notations: 1 - Earth, 2 - space tower about 100 km of height, 3 - conducting *E* layer of Earth's ionosphere, 4 - back connection through Earth.

## Theory of Space Plasma Transfer for Electric Energy, Estimations and Computations

### 1. General Theory

The magnetic intensity and magnetic pressure of a electric current reaches a maximum upon the surface of a plasma cable. Let us attempt to equate plasma gas pressure to a magnetic pressure and find the requested equilibrium electric current for a given (same) temperature of electrons and ions

$$P_g = 2nkT_k, \quad P_m = \frac{\mu_0 H^2}{2}, \quad H = \frac{I}{2\pi r},$$

$$P_m = P_g, \quad I = 4\pi r \left(\frac{knT_r}{\mu_0}\right)^{0.5}, \quad T_k = \frac{m_e u_r^2}{2k}, \quad (1)$$

where $P_g$ is plasma gas pressure, N/m$^2$; $P_m$ is magnetic pressure, N/m$^2$; $n$ is plasma density (number of electron equals number of ions: $n = n_e = n_i$), 1/m$^3$; $k = 1.38 \times 10^{-23}$ is Boltzmann coefficient, J/K; $\mu_0 = 4\pi 10^{-7}$ is magnetic constant, H/m; $H$ is magnetic intensity, A/m; $I$ is electric currency, A; $r$ is radius of plasma cable, m; $T_r$ is plasma temperature in radial direction of plasma cable, K; $m_e = 9.11 \times 10^{-31}$ is electron mass, kg; $u_r$ is electron speed in radial direction of plasma cable, m/s.

From (1) we receive relation between a minimal electric current $I_{min}$, gas density $n$ and the radial temperature of electrons

$$I_m = 4\pi r \left(\frac{knT_r}{\mu_0}\right)^{0.5} \approx 4.16 \times 10^{-8} r\sqrt{nT_r},$$

$$j_m = \frac{I}{\pi r^2} = 4\left(\frac{k}{\mu_0}\right)^{0.5} \frac{\sqrt{nT_r}}{r} \approx 1.33 \cdot 10^{-8} \frac{\sqrt{nT_r}}{r}, \quad (2)$$

where $I_m$ is minimal electric current, A; $j_m$ is density of electric current, A/m$^2$; $\pi r^2 = S$ is the cross-section area of plasma cable, m$^2$.

Assume the temperature (energy) of electrons equals temperature (energy) of ions. Let us to write well-known relations

$$j = en(u_i + u_e), \quad \frac{m_i u_i^2}{2} = \frac{m_e u_e^2}{2} \quad (3)$$

where $e = 1.6 \times 10^{-19}$ C is charge of electron, C; $m_e = 9.11 \times 10^{-31}$ kg is mass of electron, kg; $m_i$ is mass of ion, kg (for $H_2$ $m_i = 2 \times 1.67 \times 10^{-27}$ kg); $u_i$, $u_e$ is speeds of ions and electrons respectively **along** cable axis produced by electric intensity (electric generator), m/s.

The computation $j$ by Eq. (2) is presented in Figure 9.

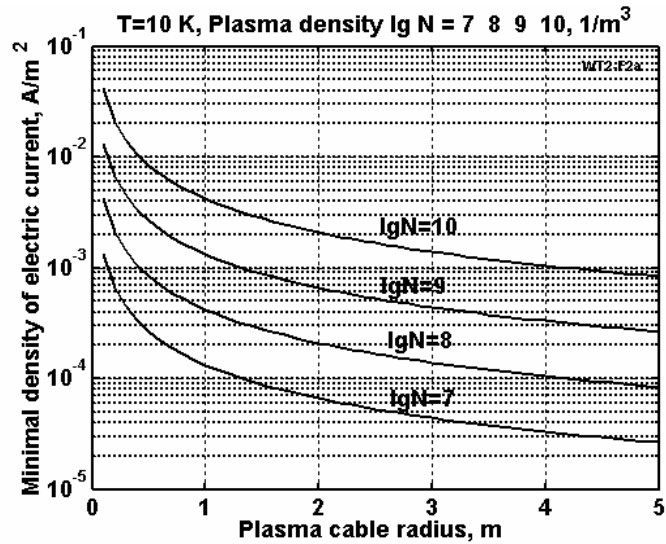

**Figure 9**. Minimal density of electric current in plasma cable for radial plasma temperature $10^o$K.

From (3) we receive axial speeds of ions and electrons produced by electric intensity (electric generator)

$$u_e = \frac{j}{en(1+\sqrt{m_e/m_i})}, \quad u_i = \frac{j}{en(1+\sqrt{m_i/m_e})} \quad . \tag{4}$$

or

$$u_e \approx 6{,}15 \cdot 10^{18} \, j/n \quad \text{for } H_2 \quad u_i \approx 10^{17} \, j/n, \quad u_e \gg u_i, \quad u = u_e + u_i \approx u_e. \tag{4'}$$

or $u \approx j/en$.

Under electric intensity the electrons and ions have opposed speeds along cable axis. The computation of electron speed produced by electric current is presented in Fig.10.

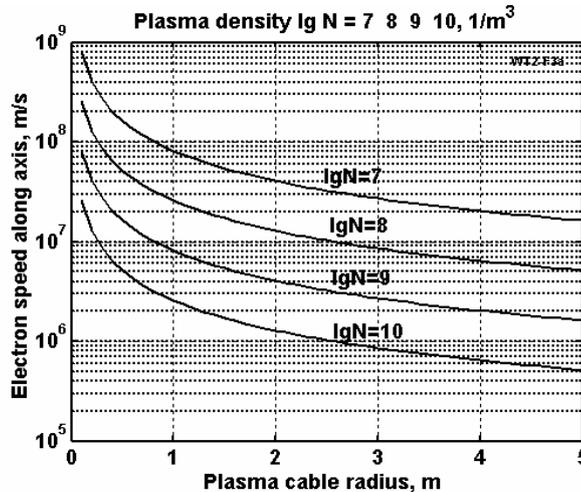

**Figure 10**. The electron speed produced by the electric current of the minimal current density versus plasma cable radius. The ions ($H_2$) speed is less 61.5 times and opposed the electron speed.

Temperature induces by electric voltage is

$$T_k = \frac{m_e u^2}{2k} \approx 3.3 \cdot 10^{-8} u^2 \ [K], \quad T = \frac{k}{e} T_k \approx 2.71 \cdot 10^{-12} u^2 \ [eV], \quad (5)$$

where $T_k$ is induced temperature in K; $T$ is this temperature along cable axis in eV. Computation is shown in Fig.11.

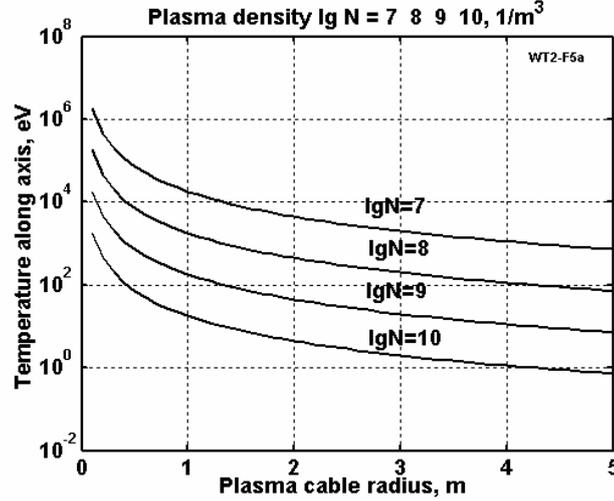

**Figure 11**. The temperature of electron and ions ($H_2$) (in eV) produced by the electric current in the minimal current density versus the plasma cable radius and different plasma density. It is assumed the ions ($H_2$) temperature equals the electron temperature.

Specific Spitzer plasma resistance and typical resistance of a plasma cable can be computed by equations:

$$\rho = \eta_\perp = 1.03 \times 10^{-4} Z \ln \Lambda T^{-3/2} \quad \Omega \cdot m, \quad R = \rho L / S, \quad (6)$$

where $\rho$ is specific plasma resistance, $\Omega \cdot m$; $Z$ is ion charge state, $\ln \Lambda \approx 5 \div 15 \approx 10$ is the Coulomb logarithm; $T = T_k k/e = 0.87 \times 10^{-4} T_k$ is plasma temperature along cable axis in eV; $e = 1.6 \times 10^{-19}$ is electron charge, C; $R$ is electric resistance of plasma cable, $\Omega$; $L$ is plasma cable length, m; $S$ is the cross-section area of the plasma cable, $m^2$.

The computation of the specific resistance of a plasma cable is presented in Figure 12.

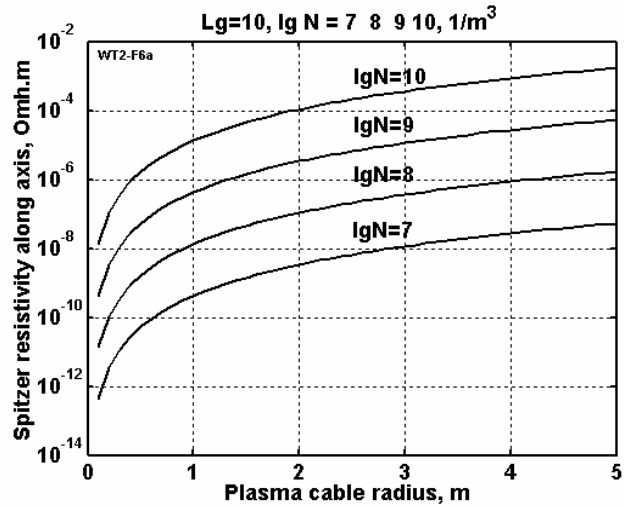

**Figure 12.** Specific (Spitzer) plasma resistance $\Omega \cdot m$ of equilibrium plasma cable for the minimal electric current versus cable radius and different plasma density. Coulomb logarithm equals 10.

The requested minimum voltage, power, transmitter power and coefficient of electric efficiency are:
$$U_m = IR, \quad W_m = IU_m, \quad U = U_m + \Delta U, \quad W = IU, \quad \eta = 1 - W_m/W = 1 - U_m/U \qquad (5)$$
where $U_m$, $W_m$ are requested minimal voltage, [V], and power, [W], respectively; $U$ is used voltage, V; $\Delta U$ is electric voltage over minimum voltage, V; $W$ is used electric power, W; $\eta$ is coefficient efficiency of the electric line. If $\Delta U >> U_m$ the coefficient efficiency closed to 1.

Computation of loss voltage and power into plasma cable having length 100 million km is in Figs. 13-14.

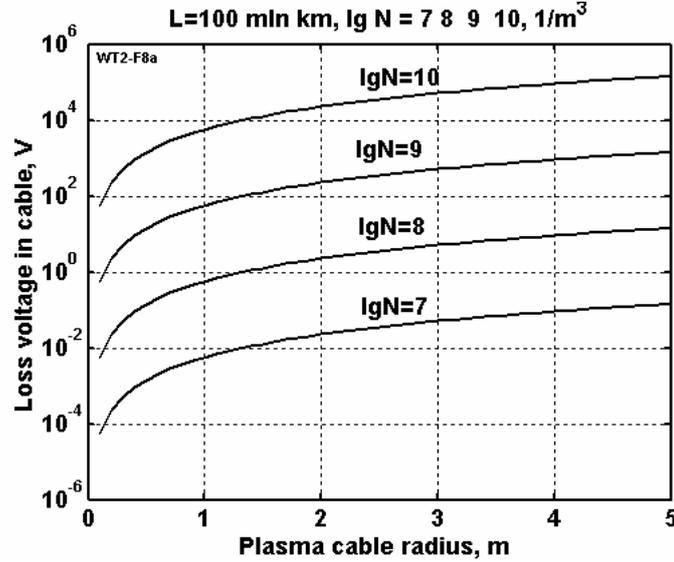

**Figure 13**. Loss voltage in plasma cable of 100 millions km length via cable radius for the minimal electric current and different plasma density.

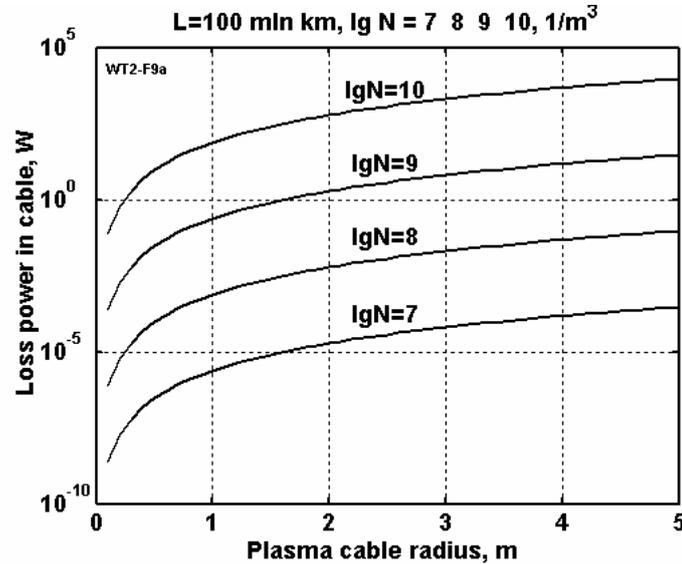

**Figure 14**. Loss power in plasma cable of 100 millions km length via cable radius for the minimal electric current and different plasma density.

The equilibrium mass $M$ [kg] of plasma cable is
$$M = \pi r^2 n m_i L, \qquad (6)$$
where $m_i$ is ion mass of plasma, kg; $L$ is length of plasma cable, m.

The mass of plasma cable is very small, about some grams for 100 millions km. The mass of a plasma cable is close to zero for any practical case when R < 5 m.

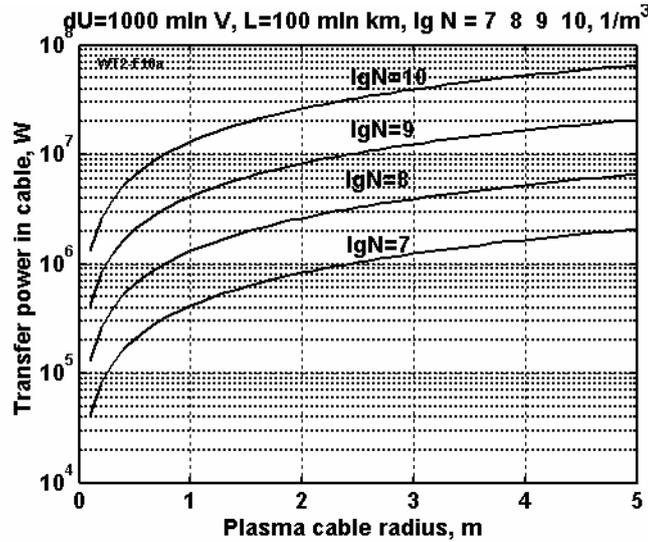

**Figure 15**. Electric power transfers by plasma cable of 100 millions km length via cable radius for the minimal electric current, over voltage $10^9$ V and the different plasma density. Coefficient efficiency is about 0.9999.

## 2. Circular Plasma Cable

The force acting in a particle (proton) moved in electric and magnetic fields may be computed by the equations:

$$\overline{F}_1 = \frac{m_i v^2}{r}, \quad \overline{F}_2 = e\overline{v}\overline{B}, \quad \overline{F}_3 = \frac{eQ_0}{4\pi\varepsilon_0 R^2}, \quad \overline{F}_4 = \gamma \frac{m_1 m_2}{R^2}, \quad \overline{F}_4 = gm_i, \qquad (7)$$

where $F_1$, $F_2$, $F_3$, $F_4$ are centrifugal, Lorenz, electrostatic, and gravitational forces respectively (all vectors), N; $m_p = 1.67 \times 10^{-27}$ kg mass of proton (or ion $m_i$); $v$ - speed of particle, m/s; $e$ - electron (proton) charge; $B$ - total magnetic induction (magnetic field strength), T; $Q_0$ - charge of central body, C; $\varepsilon_0 = 8.85 \times 10^{-12}$ F/m - electric constant; $m_1$, $m_2$ are mass of bodies (central and particle), kg; $\gamma$ – gravitational constant (for Earth $\gamma = 6.67 \times 10^{-11}$ m³/kg·s², $g_o = 9.81$ m/s²; for Sun $g_o = 274$ m/s²); $r$ – radius curve, m; $R$ – distance between charges (bodies), m.

The equilibrium condition is:

$$\sum_i F_i = 0 \qquad (8)$$

## 3. Magnetic Pressure from the Plasma Cable

The plasma exerts a pressure within the plasma cable. This pressure is small, but the cable can has a large diameter (up 200 m or more) and this pressure acting over a long time can accelerate or brake a space apparatus with no reaction mass. This magnetic pressure $P$ [N/m²] can be computed by equations:

$$P_m = \frac{\mu_0 H^2}{2}, \quad H = \frac{I}{2\pi r}, \quad P = \frac{1}{2} 2P_m S = \frac{\mu_0}{4\pi} I^2, \qquad (9)$$

*Estimation*. For $I = 10^4$ A, the magnetic pressure equals 10 N, for $I = 10^5$ A, it equals 1000 N.

## 4. Electric Pressure from the Plasma Cable.

The high speed electrons and ions of electric current within plasma cable have kinetic energy. This energy produces an electric pressure when space ship or final station uses the electric energy. Let as to estimate the electric pressure.

Specific (kinetic) energy of electric current into plasma cable is

$$E = 0.5n(m_e u_e^2 + m_i u_i^2), \ [W/m^3] \tag{10}$$

Substitute Eqs.(4) in (10) we have

$$E = \frac{j^2}{2ne^2}\left[\frac{m_e}{(1+\sqrt{m_e/m_i})^2} + \frac{m_i}{(1+\sqrt{m_i/m_e})^2}\right] \approx \frac{j^2}{2ne^2}[m_e + m_e] = \frac{m_e}{e^2}\frac{j^2}{n}. \tag{11}$$

But specific energy equals the specific pressure $P = E$ [N/m²].

$$P = \frac{m_e}{e^2}\frac{j^2}{n} \approx 3.36 \times 10^7 \frac{j^2}{n}. \tag{12}$$

*Estimation*: For $j = 100$ A/m², $n = 10^{10}$ 1/m³ we get $P = 35.6$ N/m².

## 5. Additional Power from a Space Apparatus' Motion

This power is:

$$W = PV, \tag{13}$$

where $V$ is apparatus speed, m/s.

*Estimation*. For $V = 11$ km/s, $P = 10^{-3}$ N, this power equals 11 W, for $P = 1$ N the power equals 11000 Watts. We spend this power when space apparatus moves away from the energy source ('launch point') and receive it when apparatus approaches to the energy station. ('landing site')

## 6. Track Length of Plasma Electrons and Ions

The track length $L$ and the track time $\tau$ of particles is

$$L = \upsilon_T / \nu, \ \tau = 1/\nu, \tag{14}$$

where $\upsilon_T$ is particle velocity, cm/s; $\nu$ is particle collision rate, 1/s.
The electron, ion, and electron-ion **thermal** collision rate are respectively:

$$\begin{aligned}\nu_e &= 2.91 \times 10^{-6} n_e \ln \Lambda T_e^{-3/2} \ \ s^{-1} \\ \nu_i &= 4.80 \times 10^{-8} Z^4 \mu^{-1/2} n_i \ln \Lambda T_i^{-3/2} \ \ s^{-1}, \\ \nu_{ei} &= 4.4 \times 10^{-6} n_i \lg \Lambda T^{-3/2}.\end{aligned} \tag{15}$$

where $Z$ is ion charge state, $\ln \Lambda \approx 5 \div 15 \approx 10$ is Coulomb logarithm, $\mu = m_i/m_p$ is relative mass of ion; $m_p = 1.67 \times 10^{-27}$ is mass of proton, kg; $n$ is density of electrons and ions respectively, 1/cm³; $T$ is temperature of electron and ion respectively, eV.

Electron and ion **terminal** velocity are respectively:

$$\begin{aligned}\upsilon_{Te} &= (kT_e/m_e)^{1/2} = 4.19 \times 10^7 T_e^{1/2} \ \ cm/s \\ \upsilon_{Ti} &= (kT_i/m_i)^{1/2} = 9.79 \times 10^7 \mu^{-1/2} T_i^{1/2} \ \ cm/s\end{aligned} \tag{16}$$

Substitute equations (12)-(13) in (11) we receive the length of electron and ion tracks:

$$\begin{aligned}L_e &= 1.44 \times 10^{13} T_e^2 / n_e \ln \Lambda \ \ cm, \\ L_i &= 2.04 \times 10^{13} T_e^2 / Z^4 n_e \ln \Lambda \ \ cm, \\ L_{ei} &= 0.95 \times 10^{13} T_e^2 / n_e \ln \Lambda \ \ cm.\end{aligned} \tag{17}$$

*Estimation*. For electron having $n = 10^5$ 1/cm³, $T = 100$ eV, $\ln \Lambda \approx 10$ we get $L = 2 \times 10^6$ km, $\tau \approx 300$ s.

That means the plasma electrons have very few collisions, small dispersion, (in our case) and it can have different average ELECTRON (relative to ion) temperature along the cable axis and perpendicular cable axis. It is not a surprise because the plasma can have different average temperatures of electron and ions. That also means that our assumption about the terminal and currency electron velocities being the same is very limited and the parameters of a plasma electric system will often be better, than in our computation. The plasma in our system may be very cool in a radial direction and simultaneously very hot in the axial direction. That decreases the electric current needed for plasma compression and allows a transfer of the plasma beam, energy, and thrust to a great distance.

## 7. Long Distance Wireless Transfer of Electricity on Earth

The transferring of electric energy from one continent to other continent through ionosphere and the Earth surface is described again. For this transferring we need two space towers of 100 km height, the towers must have a big conducting ball at their top end and underground (better, underwater) plates for decreasing the contact electric resistance (a good
Earth ground). The contacting ball is a large (up to 100 - 200 m diameter) inflatable gas balloon having a conductivity layer (covering, or coating).
Let us to offer the method which allows computation of the parameters and possibilities of this electric line.
The electric resistance and other values for a conductive medium can be estimated by the equations:

$$R = \frac{U}{I} = \frac{1}{2\pi a \lambda}, \quad W = IU = 2\pi a \lambda U^2, \quad E_a = \frac{U}{2a}, \qquad (18)$$

where $R$ is the electric resistance of a conductive medium, $\Omega$ (for sea water $\rho = 0.3$ $\Omega$·m); $a$ is the radius of the contacting (source and receiving sphere) balloon, m; $\lambda$ is the electric conductivity, $(\Omega\cdot m)^{-1}$; $E_a$ is electric intensity on the balloon surface, V/m.
The conductivity $\lambda$ of the $E$-layer of Earth's ionosphere as a rare ionized gas can be estimated by the equations:

$$\lambda = \frac{ne^2\tau}{m_e}, \quad \text{where} \quad \tau = \frac{L}{v}, \quad L = \frac{kT_k}{\sqrt{2}\pi r_m^2 p}, \quad v^2 = \frac{8kT_k}{\pi m_e}, \qquad (19)$$

where $n = 3.1 \times 10^9 \div 5 \times 10^{11}$ 1/m$^3$ is density of free electrons in $E$-layer of Earth's ionosphere, 1/m$^3$; $\tau$ is the time of electrons on their track, s; $L$ is the length traversed by electrons on their track, m; $v$ is the average electron velocity, in m/s; $r_m = 3.7 \times 10^{-10}$ (for hydrogen N$_2$) is diameter of gas molecule, m; $p = 3.2 \times 10^{-3}$ N/m$^2$ is gas pressure for altitude 100 km, N/m$^2$; $m_e = 9.11 \times 10^{-31}$ is mass of electrons, kg.
The transfer power and efficiency are

$$W = IU, \quad \eta = 1 - R_c / R, \qquad (20)$$

where $R_c$ is common electric resistance of conductivity medium, $\Omega$; $R$ is total resistance of the electric system, $\Omega$.
See the detailed computations in the Macro-Projects section.

**7. Earth's ionosphere as the gigantic storage of electric energy**. The Earth surface and Earth's ionosphere is gigantic spherical condenser. The electric capacitance and electric energy storied in this condenser can be estimated by equations:

$$C = \frac{4\pi\varepsilon_0}{1/R_0 - 1/(R_0 + H)} \approx 4\pi\varepsilon_0 \frac{R_0^2}{H}, \quad E = \frac{CU^2}{2}, \qquad (21)$$

where $C$ is capacity of condenser, C; $R_0 = 6.369 \times 10^6$ m is radius of Earth; $H$ is altitude of $E$-layer, m; $\varepsilon_0 = 8.85 \times 10^{-12}$ F/m is electrostatic constant; $E$ is electric energy, J.
The leakage currency is

$$i = \frac{3\pi\lambda_a R_0^2}{H} U, \quad \lambda_a = n_a e\mu, \quad R_a = \frac{H}{4\pi\lambda_a R_0^2}, \quad t = CR_a, \qquad (22)$$

where $i$ leakage currency, A; $\lambda_a$ is conductivity of Earth atmosphere, $(\Omega\cdot m)^{-1}$, $n_a$ is free electron density of atmosphere, 1/m$^3$; $\mu = 1.3 \times 10^{-4}$ (for N$_2$) is ion mobility, m$^2$/(sV); $R_a$ is Earth's atmosphere resistance, $\Omega$; $t$ is time of discharging in $e = 2.73$ times, s;

**8. Magnetic Sail**. Circle plasma cable allows creating the gigantic Magnetic Sail. This sail has drag into Solar wind, which can be used as a thrust of a space ship. The electric resistance of plasma

MagSail is small and MagSail can exist some years. That is also big good storage of electric energy. Space ship connects to MagSail by magnetic force.

The energy storage in plasma ring is

$$E_R = \frac{L_R I^2}{2}, \quad \text{where} \quad L_R = \mu_0 \frac{\pi R}{2}, \qquad (23)$$

where $E_R$ is energy in magnetic ring, J [15]; $L_R$ is inductance of magnetic ring, H; $R$ is radius of magnetic ring, m.

The ring spends power

$$U_m = R_m I, \quad W_m = I U_m, \qquad (24)$$

The existing time is

$$\tau \approx c_R \frac{E_R}{W_m}, \qquad (25)$$

where $c_R$ is part of ring energy spent in life time, s ($0 < c_R < 1$).
The ring energy is enough for some years of ring existing.
See the estimations in Projects section.

## Macroprojects

The macroprojects discussed below are not optimal. These are only examples of estimations: what parameters of system we can have.

1. **Space electric line the length in 100 millions of km**.
Let us take the following date of the electric line: radius of plasma cable is $r = 150$ m, (cross-section of plasma cable equals $S = \pi r^2 = 7.06 \times 10^4$ m$^2$), plasma density is $n = 10^{10}$ 1/m$^3$, electric currency is $I = 100$ A, electric voltage is $U = 2 \times 10^6$ V. Use the equations (1)-(6) we are receiving:

Electron velocity is $u = I/enS = 8.85 \times 10^5$ m/s, electron temperature in eV is $T = 2.23$ eV, electron temperature in K is $T_k = 2.59 \times 10^4$ K, specific electric resistance is $\rho = 3 \times 10^{-4}$ $\Omega$·m, Coulomb logarithm is $\ln \Lambda = 10$, charge state is $Z = 1$, electric resistance is $R = 2\rho L/S = 8.8 \times 10^2$ $\Omega$, loss voltage is $U_m = IR = 8.8 \times 10^4$ V, loss power is $W_m = IU_m = 8.8 \times 10^6$ W, transfer power is $W = IU = 2 \times 10^8$ W, coefficient efficiency is $\eta = 0.956$.

As you see, our system can transmit 200 million watts of power at distance 100 million kilometers with efficiency 95.6%. I remind that the distance to Mars is only about 60 million of kilometers.
Mass of our plasma line from hydrogen H$_2$ is only 470 g.

**2. Transferring electric energy to Moon or back**.
Let us take the initial data: radius of plasma cable $r = 15$ m ($S = \pi r^2 = 706$ m$^2$), plasma density $n = 10^{12}$ 1/m$^3$, electric currency $I = 1000$ A, distance 385,000 km.

Then: $u = I/enS = 8.85 \times 10^6$ m/s, $T = 223$ eV, $T_k = 2.59 \times 10^6$ K, $\rho = 3.1 \times 10^{-7}$ $\Omega$·m, $\ln \Lambda = 10$, $Z = 1$, $R = 2\rho L/S = 3.4 \times 10^{-1}$ $\Omega$, $U_m = IR = 3.4 \times 10^2$ V, $W_m = IU_m = 3.4 \times 10^5$ W.

If voltage is $U = 3.4 \times 10^3$ V, then transmitting power is $W = IU = 3.4 \times 10^8$ W, coefficient efficiency is $\eta = 0.9$.

If $U = 3.4 \times 10^4$ V, then $W = IU = 34 \times 10^8$ W, $\eta = 0.99$.

As you see, this system can transmit 340 ÷ 3400 million watts of power to Moon at distance 385,000 kilometers with efficiency 90 ÷ 99%.

If we take electric currency $I = 100$ A and voltage $U = 3.4 \times 10^3$ V, then the transfer energy is $W = IU = 3.4 \times 10^7$ W, $\eta = 0.9$. The same parameters are transfer energy to Earth's Space Station. Now the International Space Station has only electric power W = $10^4$ W.

**3. Transferring energy to Mars** located beyond the in Sun opposed on Earth side. In this case we use the circle plasma transfer (Fig. 5).

Let us take the following initial data: Radius of circle $R = 1.9 \times 10^{11}$ m = 190 millions kilometers (Length of circle equals $L \approx 12 \times 10^{11}$ m), $r = 150$ m ($S = \pi r^2 = 7.06 \times 10^4$ m$^2$), $n = 10^{10}$ 1/m$^3$, $I = 100$ A, $U = 10^7$ V.

Then: $u = I/enS = 8.85 \times 10^5$ m/s, $T = 2.23$ eV, $T_k = 2.59 \times 10^4$ K, $\rho = 3.1 \times 10^{-4}$ $\Omega$·m, ln $\Lambda = 10$, $Z = 1$, $R = \rho L/S = 5.27 \times 10^3$ $\Omega$, $U_m = IR = 5.27 \times 10^5$ V, $W_m = IU_m = 5.27 \times 10^7$ W, $W = IU = 2 \times 10^8$ W, $\eta \approx 0.95$.

Mass of our plasma line from hydrogen $H_2$ is only about 3 kg.

**4. Plasma Magnetic Sail** (Fig. 6). Let us take the following initial data: radius of MagSail $R = 5 \times 10^4$ m = 50 km, $r = 1.5 \times 10^3$ m ($S = \pi r^2 = 7.06 \times 10^6$ m$^2$), $n = 10^8$ 1/m$^3$, $I = 10^4$ A.

Then: $u = I/enS = 8.85 \times 10^7$ m/s, $T = 2.23 \times 10^4$ eV, $T_k = 2.59 \times 10^8$ K, $\rho = 3.1 \times 10^{-10}$ $\Omega$·m, ln $\Lambda = 10$, $Z = 1$, $R_m = \rho L/S = 1.38 \times 10^{-11}$ $\Omega$, $U_m = IR = 1.38 \times 10^{-7}$ V, $W_m = IU_m = 1.4 \times 10^{-3}$ W.

If $U = 100$ V, the ring energy is $E = 5 \times 10^6$ J [15]. If we spent only 10% of the ring energy, our MagSail will work about 10 years.

The gigantic plasma space MagSail is also an excellent storige of electric energy. If we take $U = 10^5$ V, the ring will keep about $E = 5 \times 10^9$ J.

**5. Wireless transferring energy between Earth's continents** (Fig. 7). Let us take the following initial data: Gas pressure at altitude 100 km is $p = 3.2 \times 10^{-3}$ N/m$^2$, temperature is 209 K, diameter nitrogen $N_2$ molecule is $3.7 \times 10^{-10}$ m, the ion/electron density in ionosphere is $n = 10^{10}$ 1/m$^3$, radius of the conductivity inflatable balloon at top the space tower (mast) is $a = 100$ m (contact area is $S = 1.3 \times 10^5$ m$^2$), specific electric resistance of a sea water is 0.3 $\Omega$·m, area of the contact sea plate is $1.3 \times 10^3$ m$^2$.

The computation used equation (15)-(19) give: electron track in ionosphere is $L = 1.5$ m, electron velocity $\upsilon = 9 \times 10^4$ m/s, track time $\tau = 1.67 \times 10^{-5}$ s, specific resistance of ionosphere is $\rho = 4.68 \times 10^{-3}$ $(\Omega \cdot m)^{-1}$, contact resistance of top ball (balloon) is $R_1 = 0.34$ $\Omega$, contact resistance of the lower sea plates is $R_2 = 4.8 \times 10^{-3}$ $\Omega$, electric intensity on ball surface is $5 \times 10^4$ V/m.

If the voltage is $U = 10^7$ V, total resistance of electric system is $R = 100$ $\Omega$, then electric currency is $I = 10^5$ A, transferring power is $W = IU = 10^{12}$ W, coefficient efficiency is 99.66%. In practice we are not limited in transferring any energy in any Earth's point having the 100 km space mast and further transfer by ground-based electric lines in any geographical region of radius 1000 ÷ 2000 km.

**6. Earth's ionosphere as the storage electric energy**. It is using the equations (18)-(19) we find the Earth's-ionosphere capacity $C = 4.5 \times 10^{-2}$ C. If $U = 10^8$ V, the storage energy is $E = 0.5CU^2 = 2.25 \times 10^{14}$ J. That is large energy.

Let us now estimate the leakage of current. Cosmic rays and Earth's radioactivity create 1.5 ÷ 10.4 ions every second in 1 cm$^3$. But they quickly recombine in neutral molecule and the ions concentration is small. We take the ion concentration of lower atmosphere $n = 10^6$ 1/m$^3$. Then the specific conductivity of Earth's atmosphere is $2.1 \times 10^{-17}$ $(\Omega \cdot m)^{-1}$. The leakage currency is $i = 10^{-7} \times U$. The altitude of $E$-layer is 100 km. We take a thickness of atmosphere only 10 km. Then the conductivity of Earth's atmosphere is $10^{-24}$ $(\Omega \cdot m)^{-1}$, resistance is $R_a = 10^{24}$ $\Omega$, the leakage time (decreasing of energy in $e = 2.73$ times) is $1.5 \times 10^5$ years.

As you can clearly see the Earth's ionosphere may become a gigantic storage site of electricity.

## Discussion

The offered ideas and innovations may create a jump in space and energy industries. Author has made initial base researches that conclusively show the big industrial possibilities offered by the methods and installations proposed. Further research and testing are necessary. As that is in science, the obstacles can slow, even stop, applications of these revolutionary innovations. For example, the plasma cable may be unstable. The instability mega-problem of a plasma cable was found in tokomak R&D, but it is successfully solved at the present time. The same method (rotation of plasma cable) can be applied in our case.

## Summary

This new revolutionary idea - wireless transferring of electric energy in the hard vacuum of outer space is offered and researched. A rare plasma power cord as electric cable (wire) is used for it. It is shown that a certain minimal electric currency creates a compressed force that supports the plasma cable in the compacted form. Large amounts of energy can be transferred hundreds of millions of kilometers by this method. The requisite mass of plasma cable is merely hundreds of grams. It is computed that the macroprojects: transferring of hundreds of kilowatts of energy to Earth's International Space Station, transfer energy to Moon or back, transferring energy to a spaceship at distance of hundreds of millions of kilometers, transfer energy to Mars when it is on the other side of the Sun wirelessly. The transfer of colossal energy from one continent to another continent (for example, Europe to USA and back), using the Earth's ionosphere as a gigantic storage of electric energy, using the plasma ring as huge MagSail for moving of spaceships. It is also shown that electric currency in plasma cord can accelerate or slow various kinds of outer space apparatus.

## Acknowledgement

The author wishes to acknowledge R.B. Cathcart for advice.